# The Misfit Strain Critical Point in the 3D Phase Diagrams of Cuprates

Nicola Poccia, Michela Fratini

*Department of Physics, Sapienza University of Rome, P. Aldo Moro 2, 00185 Roma, Italy*

E-mail: nicola.poccia@roma1.infn.it

## Abstract

At the time of writing, data have been reported on several hundred different cuprates materials, of which a substantial fraction show superconductivity at temperatures as high as 130 K. The existence of several competing phases with comparable energy shows up in different ways in different materials, therefore it has not been possible to converge toward a universal theory for high $T_c$ superconductivity. With the aim to find a unified description the Aeppli-Bianconi 3D phase diagram of cuprates has been proposed where the superlattice misfit strain $\eta$ is the third variable beyond doping $\delta$ and temperature T. The 3D phase diagrams for the magnetic order, and for the superconducting order extended to all cuprates families are described. We propose a formula able to describe the $T_c(\delta,\eta)$ surface, this permits to identify the stripe quantum critical point at $\delta_c=1/8$ and $\eta_c=7\%$ which is associated with the incommensurate to commensurate stripe phase transition, controlled by the misfit strain.

# 1. Introduction

The mechanism driving the emergence of a quantum macroscopic coherent phase that is able to resist to the de-coherence effects of high temperature remains a major topic of research in condensed matter. The realization of this macroscopic quantum phase in doped cuprates close to the Mott insulator regime has stimulated a large amount of investigations on the physics of strongly correlated metals. There is growing agreement that the solution of the problem of high-$T_c$ superconductivity requires the correct description of the normal state where spin, charge, orbital and lattice degree of freedoms compete and the functional phase emerges in a complex system with two main components showing mesoscopic phase separation [1]. The search for the mechanism of high $T_c$ superconductivity has been recently focusing on the identification of the critical point of a quantum phase transition [2]. These systems appear to have the common characteristic of a quantum criticality in the phase diagram, which is not present in standard low temperature BCS superconductors [3]. Peculiar behaviour at finite temperature are supposed to be caused by a quantum critical point [4]. Quantum phase transitions have been identified in different systems going from magnetic materials [5], heavy fermions [6] and ultra cold gases in optical lattice [7]. The macroscopic phase transition in the ground state of a many-body system occurs when the relative strength of two competing energy terms is varied across a critical value of a coupling term. For a superfluid system at zero temperature, by tuning a generic coupling at a critical value $g_c$, a quantum critical point appears where the superfluid long range order competes with a second different long range order [8].

There are about two hundred cuprates materials and it has been result difficult to find an universal description. Different quantum criticalities reachable by changing the doping axis have been proposed. The quantum critical point where the antiferromagnetic order competes with the superconducting order at the insulator-to-metal transition at doping $\delta=7\%$ and the quantum critical point at doping $\delta=18\%$ for the transition from the charge density wave order to disorder phase transition. Aeppli [8] and Bianconi [9-12] have proposed a different QCP introducing a new axis:

the misfit strain between layers (i.e., the internal chemical pressure) in a superlattice. At the Aeppli-Bianconi QCP the commensurate magnetic and charge order at a critical misfit strain $\eta_c$ and constant 1/8 doping [7] competes with superconducting order parameter. The superconducting high $T_c$ phase emerges in a region of incommensurate quantum magnetic [8] and lattice [9] fluctuations. In this phase bubbles of striped magnetic matter coexists with superconducting matter forming an inhomogeneous phase called "superstripes" [14].

Aeppli et al. in 1997 [8] have found evidence for quantum magnetic fluctuations in a 3D phase diagram where the magnetic order depends on the tolerance factor t between the La-O layer and the $CuO_2$ layer and the hole doping in La214 cuprate family. We have implemented their phase diagram to all cuprate families replacing the tolerance factor with the $CuO_2$ layer microstrain $\varepsilon$ [9] that measures the compression of the average lattice units in the ab plane. This quantity can be measured in all cuprate families and it is related with the superlattice misfit strain or internal chemical pressure $\eta=2\varepsilon=1-t$.

A large amount of data clearly indicates that there is a phase separation regime where superstripes bubbles appear that is not only a function of doping but also of the misfit strain acting on the $CuO_2$ layers, due to interlayer mismatch. The chemical pressure is a well established physical variable that controls the physical properties of perovskites and it is usually measured by the average ionic radius of the cations in the intercalated layers or the tolerance factor t, in fact, the internal chemical pressure in perovskites. In all perovkites and particularly in manganites [1], it is well established that the phase diagram of the electronic phases depends on the two variables, charge density and chemical pressure. Since the early years of high-$T_c$ superconductivity research the mismatch chemical pressure has been considered as a key variable controlling the electronic properties of cuprates only on one family, La214, however it was not possible to extend this idea to other families for the presence of a plurality of intercalated layers with cations having largely different coordination numbers. Therefore, it was not possible to compare the average ionic size $<r_A>$ in the intercalated layers and to get the tolerance factor t for all cuprates families. This problem was solved by obtaining the internal chemical pressure from the measure of the compressive microstrain $\varepsilon = (R_0-r)/r$ in the $CuO_2$ plane (that has the same absolute value as the tensile microstrain in the intercalated layers) where r is the average Cu-O distance and $R_0 = 197$ pm is the unrelaxed Cu-O distance.

In this way it is possible to have an universal phase diagram of cuprates and to reach the proximity of the quantum critical point at 1/8 varying, via ionic substitution, the internal chemical pressure [12]. In this paper we will describe the 3D phase diagrams for the magnetic order and for the superconducting order extended to all cuprates families.

## 2. Results

In the new phase diagram for the magnetic order parameter is plotted in Fig. 1. It is a function of chemical pressure $y=\eta/\eta_c$ normalized at the critical value $\eta_c=7\%$ and the doping $x==\delta/\delta_c$ normalized at the critical value $\delta_c=1/8$. The Aeppli-Bianconi superstripes critical point occurs at the point (1,1) in the xy plane. At this point commensurate magnetic stripes appear. There are three regions of phase separation (PS1, PS2, PS3) qualitatively different by experimental data [15, 16]. In particular from the analysis of magnetic neutron scattering experiments, there is an agreement for the frustrated mesoscopic phase separation at doping larger than 1/8 in Sr-doped La214, Y123, and Bi2212 between a first more delocalized component that does not show spin fluctuations and a second more localized electronic component [15,16], providing compelling evidence for the two-component scenario and mesoscopic phase separation at high doping $0.12<\delta<0.3$ in cuprates [14, 17-22].

In 2000, the variety of cuprates families have been unified in a unique description of the values of superconducting transition temperature $T_c$ as function of chemical internal pressure ($2\varepsilon$) and doping (holes number per Cu site) [9]. In Fig. 2 we plot the 3D surface of the superconducting critical temperature as a function of doping and misfit strain.

$$T_c(x, y) = A \cdot \frac{T_{c-\max}}{(x_b - x_a)} (x - x_a)(x_b - x) \cdot \left| \frac{1 + a\left(\frac{y - y_0}{\Gamma}\right)^2}{1 + \left(\frac{y - y_0}{\Gamma}\right)^2} \right| - Be^{-\left[\frac{(x-x_c)^2 + (y-y_c)^2}{2\sigma^2}\right]}$$

The superconducting critical temperature $T_c$ is a function of chemical pressure $y=\eta/\eta_c$ normalized at the critical value $\eta_c$=7% and the doping $x==\delta/\delta_c$ normalized at the critical value $\delta_c$=1/8. $T_c(x)$ is the well known parabola with zeros at doping $x_a$ and $x_b$ and the experimental data at constant doping $T_c(y)$ are described by an asymmetric Lorentzian which the asymmetry is quantified by the "a" value and the width by $\Gamma$. The suppression of superconductivity has been described by an inverse gaussiam centered at the Aeppli-Bianconi critical internal chemical pressure $y_c$ and doping $x_c$. A and B are normalizing constants. The numerical data used to plot the 3D surface plotted in Fig. 2 are: $T_{c-\max}$ = 130 K; $x_a$ = 0.5; $x_b$ = 2; A = 0.92; $y_0$ = 0.4; a = 0.4; $\Gamma$= 0.3; $x_c$ =1 ; $y_c$ = 1; B= 0.02 and $2\sigma^2$ = 0.05.

This phase diagram has been recently used to describe the phase separation scenario in a two band Hubbard model by Kugel et al. [21]. A similar scenario is expected in the other families of High $T_c$ superconductors such as FeAs superconductors which all present a lamellar structure [23]. The region of phase separation (denoted as PS3 in the figures 1 and 2) described in the work of Kugel et al. [22] is in the proximity of the quantum critical point at doping 1/8 and chemical internal pressure 7% which in figures 1 and 2 is the point (1,1).

Near this point, the order that competes with the superconducting long range order is the polaron electronic crystalline phase called a Wigner polaron crystal [12]. The variation of the spin gap energy as a function of misfit strain provides a strong experimental support for this proposal [22]. This critical misfit strain point often leads to a phase transition from a weakly incommensurate state to a commensurate state [23, 24]. The appearance of the superstripes phase in high $T_c$ superconductor is clearly related with this criticality, indeed, striped phase have been proposed and observed near a commensurate incommensurate transition in different materials [25, 26].

## 3.Conclusions

In conclusion among the variables which control the phase diagram of the cuprates (temperature, pressure, magnetic field, chemical composition, etc) usually are considered only the temperature and doping plotted along the vertical and horizontal axes respectively. However this picture fails to describe the large variation of critical temperature between different cuprates materials. In this work we have proposed an analytical formula describing the complex $T_c$ variation in the cuprates in a 3D phase diagram in function of doping and internal chemical pressure. The Bianconi quantum critical point occur at doping 1/8 and superlattice misfit strain 7% . This "superstripes" quantum critical point is associated to a commensurate incommensurate structural transition where the commensurate striped phase competes with the superconducting order parameter. We are extending this approach to the recent discovered Fe-As based superconductors [27,28].

**Aknowlegements:** This work was supported by European project 517039 "Controlling Mesoscopic phase separation" (COMEPHS) (2005). We thank Stefano Agrestini, Alessandro Ricci and Antonio Bianconi for helpful discussion.

N. Poccia and M. Fratini arXiv:0812.0951

# Figure Captions:

**Figure 1:** The magnetic phase diagram of cuprates in function of $T_c$, doping and internal chemical pressure divided in three different area of phase separation experimentally observed. The pressure is measured by $y=\eta/\eta_c$ and the doping is measured by $x=\delta/\delta_c$ therefore the Aeppli-Bianconi critical point is at (1,1) where the commensurate static magnetic stripe phase occurs at $\delta_c = 1/8$ hole for Cu site and internal chemical pressure $\eta_c = 7\%$.

**Figure 2:** 3D universal phase diagram of the superconducting critical temperature of cuprates. The values of colour plot of the superconducting transition temperature $T_c$ go from 0 (black) to 135 K (white). The pressure is measured by $y=\eta/\eta_c$ and the doping is measured by $x=\delta/\delta_c$ therefore the Aeppli-Bianconi critical point is at (1,1) where the superconducting critical temperature goes to zero (black). The maximum of the critical temperature occurs at (1.4, 0.5) i.e. at doping 0.16 and misfit strain 0.35.

**N. Poccia and M. Fratini arXiv:0812.0951**

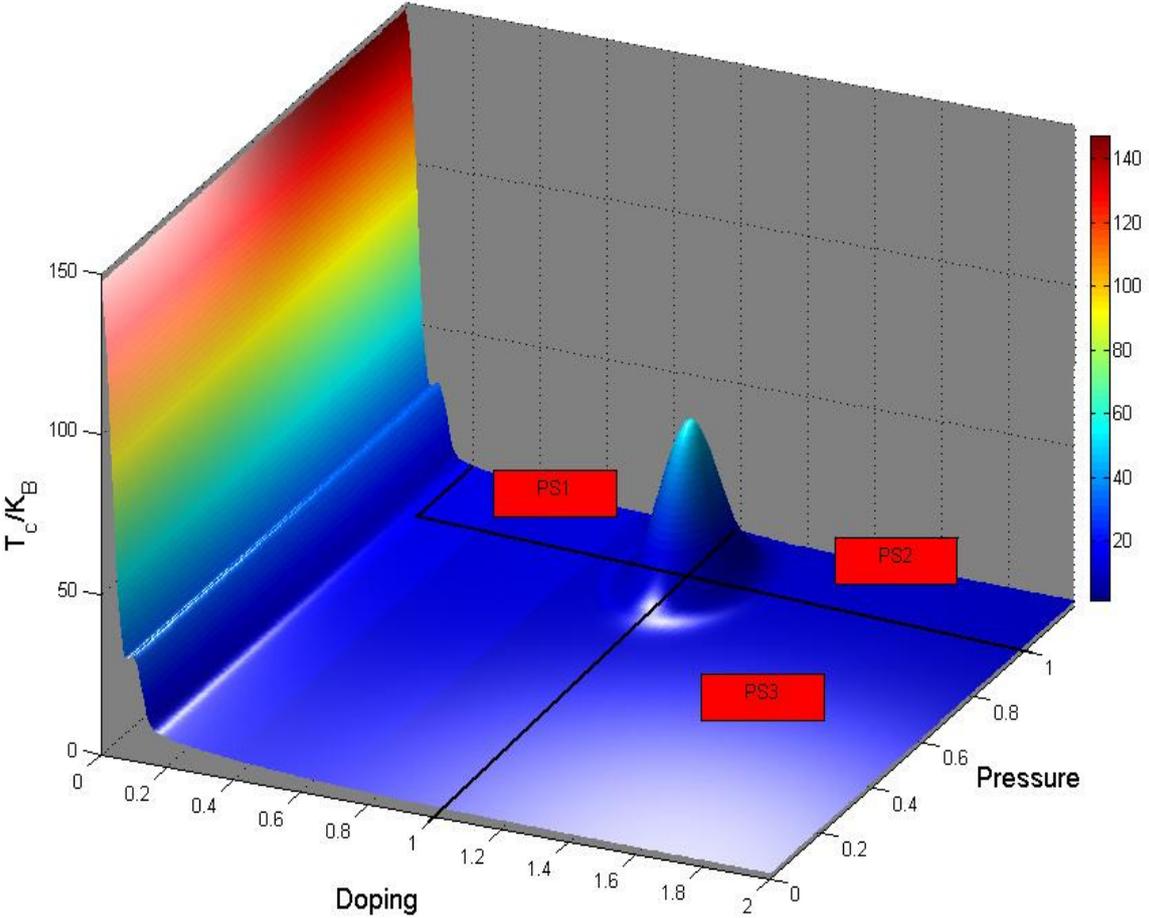

**Figure 1**

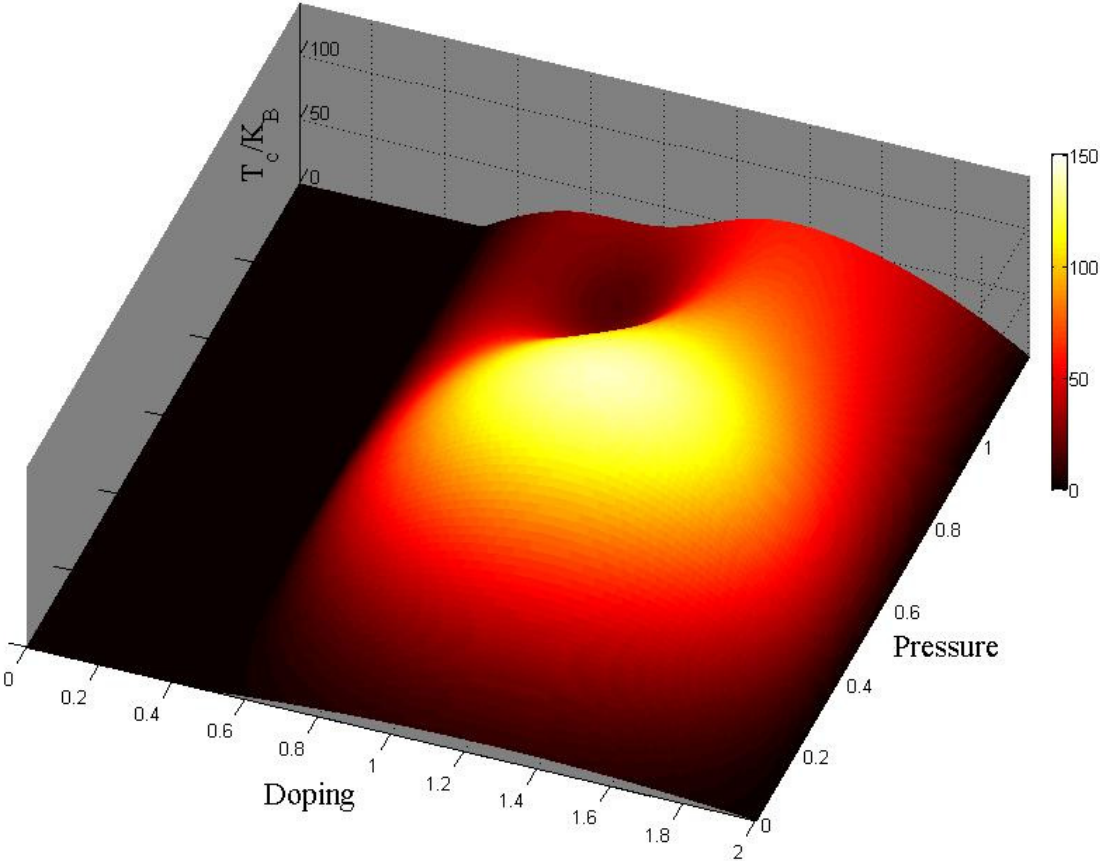

**Figure 2**